\documentclass[reprint,amsmath,amssymb,aps,prb,superscriptaddress,floatfix]{revtex4-2}

\usepackage{graphicx}
\usepackage{dcolumn}
\usepackage{bm}
\usepackage{amsfonts}
\usepackage{xcolor}
\usepackage[libertine]{newtxmath}
\usepackage{hyperref}
\usepackage{upgreek}

\hypersetup{
  colorlinks=true,
  linkcolor=blue,
  citecolor=blue,
  urlcolor=blue
}

\begin{document}

\title{Linear and nonlinear optical responses in the chiral multifold semimetal BeAu: A quantum-geometric perspective}

\author{Babu Baijnath Prasad}
\email{bbprasad@issp.u-tokyo.ac.jp}
\affiliation{Institute for Solid State Physics, The University of Tokyo, Kashiwa, Chiba 277-8581, Japan}

\author{Taisuke Ozaki}
\email{t-ozaki@issp.u-tokyo.ac.jp}
\affiliation{Institute for Solid State Physics, The University of Tokyo, Kashiwa, Chiba 277-8581, Japan}

\date{\today}

\begin{abstract}
Chiral topological semimetals provide a natural platform for exploring how multifold band topology and quantum geometry manifest in optical and photovoltaic responses. 
BeAu is a chiral multifold semimetal hosting band crossings at $\Gamma$, $M$, and $R$ with Chern numbers $C_{\Gamma}=-4$, $C_{M}=-2$, and $C_{R}=+4$, respectively. 
In this work, we study the linear optical conductivity and second-order dc photocurrent responses of BeAu using fully relativistic first-principles calculations combined with Wannier interpolation.
The calculated interband linear optical conductivity, $\mathrm{Re}\,\sigma_{xx}(\omega)$, is quantitatively reproduced by $(e^{2}/\hbar)\omega g_{xx}(\omega)$, demonstrating that its spectral features are governed jointly by the explicit photon energy factor and the variation of the photon energy-resolved quantum-metric spectral weight.
The linear shift current conductivity is closely related to the symplectic connection, whereas the circular injection current susceptibility is governed by the transition-resolved product of Berry curvature and the interband group velocity difference.
Notably, at the Fermi level, the linear shift current conductivity exhibits an exceptionally large low-energy response, reaching approximately $-810~\mu\mathrm{A/V^2}$ at a photon energy of $0.05$ eV.
Aligning the chemical potential with the multifold crossings strongly reshapes both responses, producing the largest linear shift current conductivity peak for $\mu=\mu_R$ and pronounced changes in the magnitude and sign of the circular injection current susceptibility.
The circular photogalvanic trace is strongly photon-energy and chemical-potential dependent and does not exhibit a broad quantized plateau, indicating competing multiband transitions. 
Our results thus establish a unified quantum-geometric description of the linear and nonlinear optical responses of BeAu and identify it as a promising platform for optoelectronic phenomena governed by multifold band topology and quantum geometry.
\end{abstract}

\maketitle

\section{INTRODUCTION}

Topological semimetals host band crossings around which the Bloch wave
functions acquire nontrivial geometric and topological structures. 
In a Weyl semimetal, an isolated twofold crossing acts as a monopole of Berry
curvature and carries an integer Chern number \cite{Wan2011}. 
Additional crystalline symmetries can protect higher-fold degeneracies, giving rise
to multifold fermions beyond the conventional Weyl and Dirac classification \cite{Bradlyn2016}. 
Chiral crystals are particularly favorable for realizing such quasiparticles and can host  
multifold band crossings with large topological charges. 
Prominent examples are the transition-metal monosilicides and related compounds in the chiral cubic space group $P2_{1}3$, where fourfold and sixfold fermions occur at the
time-reversal-invariant $\Gamma$ and $R$ points, respectively \cite{Tang2017,Chang2017,Schroter2019,Takane2019,Sanchez2019}.

Linear optical responses provide a sensitive probe of the low-energy electronic structure of topological semimetals.
For an ideal three-dimensional Weyl fermion, the interband optical conductivity is linear in photon frequency \cite{SanchezMartinez2019}.
Multifold fermions host a larger number of optically active bands and can therefore exhibit additional absorption thresholds and characteristic changes in the magnitude and slope of the optical conductivity \cite{SanchezMartinez2019,Habe2019}.
Previous optical studies of CoSi, RhSi, and PdGa have revealed optical responses that deviate from the simple behavior of an isolated ideal node, including quasilinear regimes and additional spectral features associated with their realistic multiband electronic structures \cite{Xu2020,Maulana2020,Ni2020,Maulana2021}.

Recent developments have shown that optical responses can provide
information about the geometry of Bloch states. 
The quantum metric and Berry curvature are associated with the symmetric and antisymmetric parts, respectively, of the quantum geometric tensor and characterize
complementary aspects of the variation of Bloch wave functions in
momentum space \cite{Provost1980,Ahn2020}. 
Since optical transitions are governed by interband matrix elements, they can provide access to these geometric quantities \cite{Ozawa2018,Ahn2020}. 
This perspective is particularly relevant to multiband semimetals, where several interband
transitions generally contribute to the observed optical response.

The absence of inversion symmetry permits second-order dc photocurrents,
commonly referred to as the bulk photovoltaic effect (BPVE)
\cite{Sturman1992,Sipe2000,Morimoto2016}. 
The shift current is associated with a displacement of the electronic charge center during an interband transition, whereas the injection current arises from an asymmetric population of photoexcited carriers with different group velocities \cite{Sipe2000,Ahn2020}. 
In a nonmagnetic crystal with time-reversal symmetry, linearly polarized light generates the conventional shift current, while circularly polarized light generates the injection current \cite{Sipe2000,Ahn2020}. 

The BPVE is also closely connected to quantum geometry. 
The circular injection current involves the Berry curvature weighted by the interband
velocity difference, whereas the linear shift current is related to the
symplectic 
connection \cite{Ahn2020,Ahn2022,Prasad2024}. 
The strong variation of these geometric quantities near band crossings can enhance the low-frequency photocurrent and may enable terahertz photodetection \cite{Ahn2020}.
For an isolated chiral node, the trace of the circular injection tensor can become quantized at a universal value proportional to its Chern number, giving rise to the quantized circular photogalvanic effect (CPGE) \cite{deJuan2017}.
This result extends to multifold semimetals, where larger topological charges can produce correspondingly larger quantized values \cite{Flicker2018,Le2020}. 
In realistic materials, however, quantization requires a finite photon-energy window in which the relevant optical transitions are allowed while competing interband transitions remain Pauli blocked. 
A large circular photocurrent therefore does not, by itself, imply a quantized CPGE.

BeAu crystallizes in the chiral cubic space group $P2_{1}3$ and is a
type-I superconductor below approximately $3.2$ K \cite{Matthias1959,Amon2018,Singh2019,Beare2019}. 
In the normal state, quantum-oscillation measurements and first-principles calculations reveal a complex multiband Fermi surface containing bands associated with unconventional chiral fermions \cite{Rebar2019}. 
More recent theoretical studies have predicted significant intrinsic spin Hall conductivity and nontrivial magnetotransport \cite{Bhowmick2025}, together with multifold crossings, additional Weyl points, nodal surfaces, topological Fermi-surface sheets, and surface Fermi arcs \cite{Vocaturo2026}.

BeAu has also recently been proposed as a favorable candidate for a
metallic electro-optic response. 
A first-principles survey of 37 experimentally reported nonmagnetic compounds in space group 198 identified BeAu as the most viable material among those considered for
the magnetoelectric electro-optic effect, with a predicted critical bias field of approximately $8\times10^{5}\ \mathrm{V\,m^{-1}}$ \cite{Ascencio2025}. 
This effect is an intraband Fermi-surface response arising from Berry curvature and the orbital magnetic moment of Bloch electrons and is distinct from the interband optical processes studied here. 
Nevertheless, it further motivates a systematic investigation of optical phenomena associated with Bloch-state geometry in BeAu.

Despite extensive studies of its superconducting, transport, and topological properties, the normal-state optical response of BeAu remains largely unexplored. 
To our knowledge, its interband optical conductivity, BPVE, and photon energy-resolved quantum-geometric quantities have not been investigated systematically. 
The optical response is expected to involve transitions among several bands because the multifold crossings at $\Gamma$, $M$, and $R$ lie at different energies, while additional
bands and Fermi-surface pockets are also present. 
It is therefore important to determine how the different multifold crossings and the surrounding electronic structure shape the optical spectra, and how the responses change when the chemical potential is aligned with each crossing.

In this paper, therefore, we present a systematic first-principles study of the linear optical conductivity and bulk photovoltaic responses of chiral BeAu. 
We consider both the Fermi level and chemical potentials aligned with the multifold crossings at $\Gamma$, $M$, and $R$, and analyse the calculated spectra in terms of photon energy-resolved quantum-geometric quantities and the topological charges of the multifold nodes. 
The rest of this paper is organized as follows. 
In Sec.~II, we present the crystal structure of BeAu together with the
theoretical framework and computational details. 
The main results are presented in Secs.~III and IV.
In Sec.~III, we discuss the linear optical conductivity and its relation to the quantum metric.
In Sec.~IV, we present the bulk photovoltaic responses and interpret them in terms of the photon energy-resolved quantum geometric quantities.
The CPGE trace and its relation to the topological charges of the multifold nodes are also discussed in this section.
Finally, the conclusions drawn from this work are summarized in Sec.~V. 
Additional details on the optical conductivity--quantum metric relation, the Wannier-interpolated band structure, and the topological charges of the multifold crossings are provided in Appendices~A--C, respectively.

\section{THEORY AND COMPUTATIONAL DETAILS}

\begin{figure}[t]
\centering
\includegraphics[width=\columnwidth]{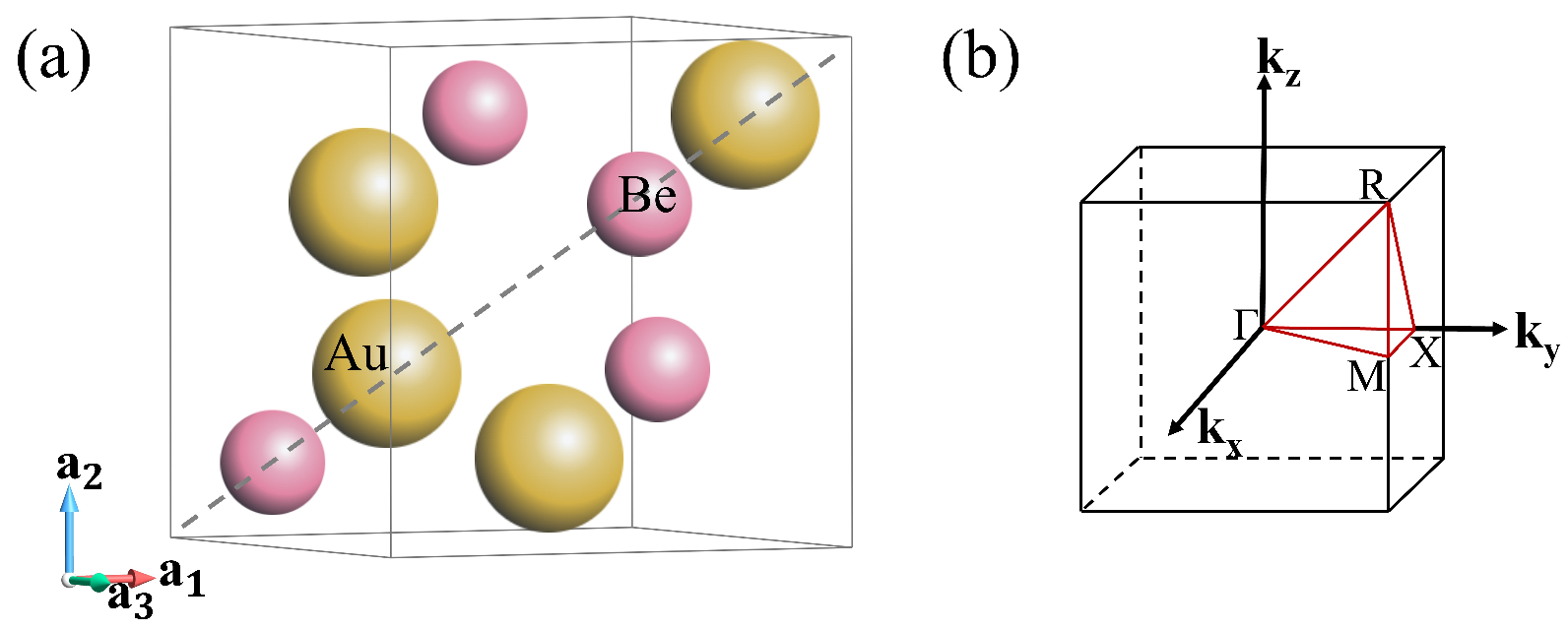}
\caption{(a) Crystal structure of chiral cubic BeAu, generated using OpenMX Viewer~\cite{OpenMXViewer}, with the dashed gray line indicating the threefold rotation axis along the [111] direction. (b) The corresponding cubic Brillouin zone.}
\label{fig:BeAu-crystal}
\end{figure}

BeAu crystallizes in the chiral cubic space group $P2_{1}3$ (No.~198), with crystallographic point group $23$. 
The primitive unit cell contains four Be and four Au atoms, both occupying the $4a$ Wyckoff positions. 
In the present calculations, we use the experimental lattice constant $a=4.668$~\AA{} \cite{Cullity1947} along with internal coordinates $u_{\mathrm{Be}}=0.150$ and $u_{\mathrm{Au}}=0.844$.
The crystal structure and the corresponding bulk Brillouin zone are shown in Figs.~\ref{fig:BeAu-crystal}(a) and \ref{fig:BeAu-crystal}(b), respectively.

The electronic-structure calculations are performed within density functional theory using the OpenMX code \cite{Ozaki2003,Ozaki2004,Ozaki2005,OpenMX}. 
The exchange-correlation energy is treated within the generalized-gradient approximation of Perdew, Burke, and Ernzerhof \cite{Perdew1996}. 
The Kohn--Sham wave functions are expanded in finite-range pseudo-atomic orbitals, and norm-conserving pseudopotentials from the OpenMX database Ver.~2019 are employed. 
The basis sets are Be7.0-$s3p2d1$ and Au7.0-$s3p3d2f1$, where the numerical value specifies the cutoff radius of the pseudo-atomic orbitals in bohr, while $s3p2d1$ and $s3p3d2f1$ denote the numbers of radial functions used for the corresponding angular-momentum channels. 
Fully relativistic pseudopotentials with spin--orbit coupling are used for self-consistent noncollinear calculations.
A real-space grid cutoff of 250 Ry and a $11\times11\times11$ $k$-point mesh are used for the self-consistent calculations.

Within the independent-particle approximation, the frequency-dependent
interband optical conductivity is evaluated using the Kubo--Greenwood
formalism. 
The complex linear optical conductivity is given by \cite{Yates2007}
\begin{equation}
\sigma_{ab}(\omega)
=
\frac{i e^{2}}{\hbar}
\int\frac{d^{3}k}{(2\pi)^{3}}
\sum_{nm}
f_{mn}\,
\frac{
\omega_{mn}\,
r_{nm}^{a}r_{mn}^{b}
}{
\omega_{mn}-\omega-i\eta/\hbar
},
\label{eq:optical_conductivity}
\end{equation}
where $f_{mn}=f(\varepsilon_{m\mathbf{k}})
-f(\varepsilon_{n\mathbf{k}})$ is the difference between the
Fermi--Dirac distribution functions, and
$\omega_{mn}
=(\varepsilon_{m\mathbf{k}}-\varepsilon_{n\mathbf{k}})/\hbar$,
with $\varepsilon_{m\mathbf{k}}$ denoting the energy of band $m$ at
$\mathbf{k}$. 
The quantity
$r_{nm}^{a}
=i\langle u_{n\mathbf{k}}|
\partial_{k_a}u_{m\mathbf{k}}\rangle$
for $n\neq m$ is the interband position matrix element, where
$u_{n\mathbf{k}}$ is the cell-periodic part of the Bloch state.
The positive parameter $\eta$, set to 10 meV in the present calculations, describes the broadening of the interband transitions. 
The occupation factors are evaluated in the zero-temperature limit ($T=0$~K).
The real and imaginary parts of $\sigma_{ab}(\omega)$ represent the absorptive and dispersive parts of the interband optical response, respectively.

In a noncentrosymmetric crystal, optical excitation can generate a
second-order dc photocurrent. 
Within the length-gauge formalism, the shift current conductivity and injection current susceptibility are given by \cite{Ahn2020,Prasad2024}
\begin{align}
\sigma_{abc}^{\mathrm{sh}}(\omega)
={}&
-\frac{i\pi e^{3}}{\hbar^{2}}
\int\frac{d^{3}k}{(2\pi)^{3}}
\sum_{nm}
f_{nm}
\left(
r_{nm}^{c}r_{mn;a}^{b}
-
r_{nm;a}^{c}r_{mn}^{b}
\right)
\nonumber\\
&\times
\delta(\omega_{mn}-\omega),
\label{eq:shift_current}
\\
\eta_{abc}^{\mathrm{inj}}(\omega)
={}&
-\frac{2\pi e^{3}}{\hbar^{2}}
\int\frac{d^{3}k}{(2\pi)^{3}}
\sum_{nm}
f_{nm}\,
\Delta_{mn}^{a}\,
r_{nm}^{c}r_{mn}^{b}
\delta(\omega_{mn}-\omega),
\label{eq:injection_current}
\end{align}
where
$\Delta_{mn}^{a}=v_{mm}^{a}-v_{nn}^{a}$ is the group velocity difference between bands $m$ and $n$, and $r_{mn;a}^{b}$ denotes the generalized momentum derivative of the interband position matrix element.  
In the numerical evaluation, the Dirac delta functions are represented by Gaussian functions with a fixed broadening of 10~meV.
The steady-state injection current conductivity is
$\sigma_{abc}^{\mathrm{inj}}
=\tau\eta_{abc}^{\mathrm{inj}}$, where $\tau$ is the relaxation time of the photoexcited carriers.

To understand the geometric origin of the calculated linear and nonlinear optical responses, we also evaluate the corresponding photon energy-resolved quantum-geometric quantities. 
The symplectic
connection, quantum metric, and Berry curvature are defined as \cite{Prasad2024}
\begin{align}
\widetilde{\Pi}_{abc}^{mn}(\mathbf{k})
&=
-\mathrm{Im}\!\left[
r_{nm}^{c}(\mathbf{k})
r_{mn;a}^{b}(\mathbf{k})
\right],
\label{eq:symplectic_connection}
\\
g_{bc}^{mn}(\mathbf{k})
&=
\mathrm{Re}\!\left[
r_{nm}^{c}(\mathbf{k})
r_{mn}^{b}(\mathbf{k})
\right],
\label{eq:quantum_metric}
\\
\Omega_{bc}^{mn}(\mathbf{k})
&=
-2\,\mathrm{Im}\!\left[
r_{nm}^{c}(\mathbf{k})
r_{mn}^{b}(\mathbf{k})
\right],
\label{eq:berry_curvature}
\end{align}
respectively. 
Their photon energy-resolved forms, together with the group velocity difference, are given by \cite{Prasad2024}
\begin{align}
\widetilde{\Pi}_{abc}(\omega)
&=
 \pi\int\frac{d^{3}k}{(2\pi)^{3}}
\sum_{nm}f_{nm}\,
\widetilde{\Pi}_{abc}^{mn}\,
\delta(\omega_{mn}-\omega),
\label{eq:energy_resolved_symplectic_connection}
\\
g_{bc}(\omega)
&=
\pi\int\frac{d^{3}k}{(2\pi)^{3}}
\sum_{nm}f_{nm}\,
g_{bc}^{mn}\,
\delta(\omega_{mn}-\omega),
\label{eq:energy_resolved_quantum_metric}
\\
\Omega_{bc}(\omega)
&=
\pi\int\frac{d^{3}k}{(2\pi)^{3}}
\sum_{nm}f_{nm}\,
\Omega_{bc}^{mn}\,
\delta(\omega_{mn}-\omega),
\label{eq:energy_resolved_berry_curvature}
\\
\Delta^{a}(\omega)
&=
\pi\int\frac{d^{3}k}{(2\pi)^{3}}
\sum_{nm}f_{nm}\,
\Delta_{mn}^{a}\,
\delta(\omega_{mn}-\omega).
\label{eq:energy_resolved_group_velocity_difference}
\end{align}
Furthermore, for a diagonal component, the absorptive optical conductivity 
is related to the quantum metric by
\begin{equation}
\mathrm{Re}\,\sigma_{aa}(\omega)
=
\frac{e^{2}\omega}{\hbar}\,
g_{aa}(\omega),
\label{eq:sigma_metric_relation}
\end{equation}
as shown in Appendix~\ref{app:optical_metric}.
The linear shift current is related to the symplectic connection, whereas the circular injection current depends on both the Berry curvature and the group velocity difference
\cite{Ahn2020,Prasad2024}.

Since dense $k$ meshes are required to obtain converged linear and nonlinear optical responses, we employ the efficient Wannier interpolation scheme based on closest Wannier functions (CWFs), as implemented in \textsc{OpenMX} \cite{Ozaki2024,CWFwebsite}, with entangled
bands treated through a smoothly varying energy-window function. 
A total of 80 CWFs per unit cell, derived from the Be $s$, $p$ and Au $s$, $d$ orbitals, are constructed by fitting to the \textit{ab initio} relativistic band structure. 
The calculated Wannier-interpolated band structure closely reproduces that from the \textit{ab initio} calculation, as shown in Appendix~\ref{app:cwf_bands}. 
The required matrix elements are then calculated and transferred to the \textsc{Wannier90} package \cite{Mostofi2008,Pizzi2020} for the calculation of linear and nonlinear optical responses as well as quantum-geometric quantities.
The complex interband optical conductivity, bulk photovoltaic responses, and photon energy-resolved quantum-geometric quantities are then evaluated on a dense $k$ mesh of $250\times250\times250$.
Test calculations using several different sets of $k$ meshes show that these calculated spectra converge within a few percent.
Chemical-potential shifts are treated within the rigid-band approximation.
The topological charges of the multifold crossings are determined from the evolution of the hybrid Wannier centers using \textsc{WannierTools} \cite{Wu2018}, as described in Appendix~\ref{app:topological_charges}.

\section{LINEAR OPTICAL CONDUCTIVITY}

\begin{figure*}[t]
\centering
\includegraphics[width=\textwidth]{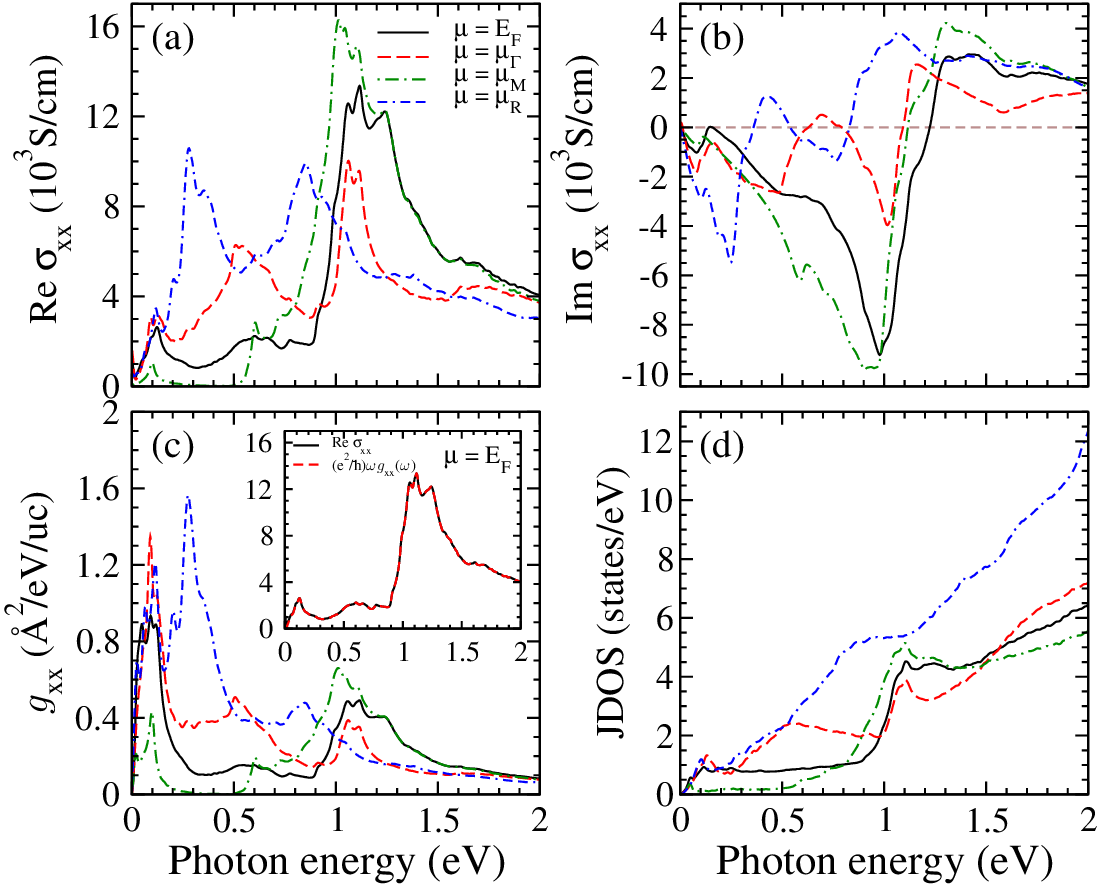}
\caption{(a) Real and (b) imaginary parts of the interband optical conductivity $\sigma_{xx}$ of BeAu at the Fermi level and at chemical potentials aligned with the multifold crossings at $\Gamma$ ($\upmu_{\Gamma}$), M ($\upmu_{\mathrm{M}}$), and R ($\upmu_{\mathrm{R}}$). (c) Corresponding photon energy-resolved quantum metric $g_{xx}(\omega)$. The inset compares the Fermi-level $\mathrm{Re}\,\sigma_{xx}(\omega)$ with $(e^{2}/\hbar)\omega g_{xx}(\omega)$. (d) Joint density of states (JDOS) for the four chemical potentials.}
\label{fig:optical}
\end{figure*}

Owing to the cubic symmetry of BeAu, the linear optical conductivity is isotropic, with $\sigma_{xx}(\omega)=\sigma_{yy}(\omega)=\sigma_{zz}(\omega)$ and vanishing off-diagonal components. 
We therefore discuss only $\sigma_{xx}(\omega)$ in the following.
The ordinary linear optical conductivity is identical for the left- and right-handed enantiomorphs.

Figures~\ref{fig:optical}(a) and \ref{fig:optical}(b) show the real and imaginary parts of the interband optical conductivity, respectively, at the Fermi level and at chemical potentials aligned with the multifold crossings at $\Gamma$, $M$, and $R$. 
At the Fermi level, $\mathrm{Re}\,\sigma_{xx}(\omega)$ rises rapidly at low photon energies and reaches a pronounced maximum at $0.12$~eV. 
It then decreases to a broad minimum at $0.32$~eV before increasing again at higher photon energies and reaching a pronounced peak near $1.10$~eV.
The corresponding imaginary part displays the dispersive response associated with these variations in the absorptive conductivity. 
In particular, $\mathrm{Im}\,\sigma_{xx}(\omega)$ remains predominantly negative over the low- and intermediate-energy range, attains a pronounced minimum at $0.98$~eV, and changes sign at approximately $1.22$~eV before reaching a broad positive maximum at $1.42$~eV [see Fig. \ref{fig:optical}(b)].

Shifting the chemical potential substantially modifies the real part of the interband optical conductivity [see Fig.~\ref{fig:optical}(a)]. 
Experimentally, such chemical-potential tuning may be achieved through controlled chemical substitution, as demonstrated in the isostructural chiral semimetal RhSi, where Ni substitution shifts the chemical potential by several hundred meV while largely preserving the underlying band structure~\cite{Sanchez2023}.
In BeAu, at $\mu=\mu_{\Gamma}$, $\mathrm{Re}\,\sigma_{xx}(\omega)$ exhibits two closely spaced low-energy maxima near $0.09$ and $0.13$~eV, followed by a broad increase from a minimum at $0.22$~eV to a
prominent maximum around $0.51$~eV. 
The dominant peak occurs at $1.06$~eV.
The response at $\mu=\mu_M$ is qualitatively different. 
After a relatively weak low-energy maximum at $0.10$~eV, the conductivity decreases to a pronounced minimum at $0.47$~eV and then rises rapidly, reaching a local maximum at $0.60$~eV and the dominant peak at $1.01$~eV. 
By contrast, at $\mu=\mu_R$, the low-energy conductivity is strongly enhanced, with the dominant maximum occurring at $0.28$~eV, followed by another broad maximum at $0.85$~eV.
The corresponding imaginary spectra are also strongly modified [see Fig.~\ref{fig:optical}(b)].  
For $\mu=\mu_{\Gamma}$, $\mathrm{Im}\,\sigma_{xx}(\omega)$ changes sign at approximately $0.63$ and $0.79$~eV, reaches a pronounced minimum at $1.02$~eV, and changes sign again at approximately $1.09$~eV before attaining a positive maximum at $1.16$~eV.
For $\mu=\mu_M$, it remains negative up to approximately $1.12$~eV before becoming positive. 
At $\mu=\mu_R$, the imaginary part changes sign at approximately $0.36$, $0.54$, and $0.83$~eV.
These pronounced changes in both $\mathrm{Re}\,\sigma_{xx}(\omega)$ and $\mathrm{Im}\,\sigma_{xx}(\omega)$ demonstrate that tuning the chemical potential modifies both the absorptive and dispersive parts of the optical response.

The quantum-geometric origin of the calculated optical spectra can be understood through the photon energy-resolved quantum metric $g_{xx}(\omega)$ shown in Fig.~\ref{fig:optical}(c). 
In the present definition, $g_{xx}(\omega)$ includes the occupation difference, the interband geometric matrix elements, and the resonant transition condition. 
It therefore represents the occupation-weighted quantum-metric spectral weight of the optically active interband transitions at each photon energy. 
As derived in Appendix~\ref{app:optical_metric}, the real part of the interband optical conductivity, $\mathrm{Re}\,\sigma_{xx}(\omega)$, is related to $g_{xx}(\omega)$ through Eq.~(\ref{eq:app_sigma_metric_relation}).
The inset of Fig.~\ref{fig:optical}(c) compares the Fermi-level $\mathrm{Re}\,\sigma_{xx}(\omega)$ with $(e^{2}/\hbar)\omega g_{xx}(\omega)$. 
The two curves overlap over the calculated photon-energy range, in agreement with Eq.~(\ref{eq:app_sigma_metric_relation}). 
This comparison shows that the frequency dependence of the absorptive conductivity is governed jointly by the explicit factor $\omega$ and the variation of the quantum-metric spectral weight with photon energy.
For example, at the Fermi level, the decrease of $\mathrm{Re}\,\sigma_{xx}(\omega)$ from the maximum at $0.12$~eV to the minimum at $0.32$~eV occurs despite the increasing photon energy and therefore reflects a pronounced reduction of $g_{xx}(\omega)$ [see Figs.~\ref{fig:optical}(a) and
\ref{fig:optical}(c)].

The chemical-potential dependence of $\mathrm{Re}\,\sigma_{xx}(\omega)$ can be understood in the same way. 
For $\mu=\mu_{\Gamma}$, the two low-energy maxima in $\mathrm{Re}\,\sigma_{xx}(\omega)$ at $0.09$ and $0.13$~eV are accompanied by a strong enhancement of the quantum-metric spectral weight. 
Its subsequent reduction is reflected in the conductivity minimum at $0.22$~eV. 
At higher photon energies, although $g_{xx}(\omega)$ is smaller than in the low-energy region, its contribution is amplified by the increasing photon-energy factor, giving rise to the conductivity maxima at $0.51$ and $1.06$~eV.
At $\mu=\mu_M$, the quantum-metric spectral weight is comparatively small in the low-energy region, and the corresponding conductivity maximum at $0.10$~eV is therefore relatively weak. 
Both quantities decrease further over the subsequent energy range, leading to a pronounced suppression of the conductivity.
The recovery of $g_{xx}(\omega)$ then produces the conductivity maximum at $0.60$~eV. 
At higher energies, $g_{xx}(\omega)$ increases substantially and becomes the largest among the four chemical-potential cases at $1.01$~eV, resulting in the dominant conductivity peak.
For $\mu=\mu_R$, the large quantum-metric spectral weight throughout the low-energy region accounts for the strongly enhanced optical conductivity. 
In particular, the pronounced peak in $g_{xx}(\omega)$ at $0.28$~eV gives rise to the dominant conductivity maximum at the same energy.
Although the quantum-metric spectral weight around $0.85$~eV is smaller than in the low-energy region, its contribution is amplified by the larger photon-energy factor, producing another prominent conductivity maximum.
Figure~\ref{fig:optical}(d) shows the joint density of states (JDOS), which characterizes the number of energetically allowed interband transitions at each photon energy.
Although some broad JDOS features occur over the same energy ranges as those in the optical spectra, the JDOS does not reproduce their relative spectral weights or detailed line shapes. 
The reduced low-energy JDOS at $\mu=\mu_M$ and its enhancement at $\mu=\mu_R$ capture the overall trends, but not the detailed energy dependence of the conductivity, highlighting the essential role of the quantum-metric spectral weight $g_{xx}(\omega)$.

These results show that the absorptive linear optical response of BeAu can be understood  
in terms of the photon energy-resolved quantum metric.
The quantitative reproduction of $\mathrm{Re}\,\sigma_{xx}(\omega)$ by $(e^{2}/\hbar)\omega g_{xx}(\omega)$ therefore provides a direct quantum-geometric interpretation of its spectral features and chemical-potential dependence.

\section{BULK PHOTOVOLTAIC RESPONSES}

The bulk photovoltaic response of nonmagnetic BeAu consists of the linear shift current and circular injection current. 
Owing to the chiral cubic point group $23$, each response has only one independent nonzero tensor component. 
For the linear shift current conductivity, $\sigma_{xyz}^{\mathrm{sh,L}} =\sigma_{yxz}^{\mathrm{sh,L}} =\sigma_{zxy}^{\mathrm{sh,L}}$,
whereas for the circular injection current susceptibility, $\eta_{xyz}^{\mathrm{inj,C}} =\eta_{yzx}^{\mathrm{inj,C}} =\eta_{zxy}^{\mathrm{inj,C}}$.
Also, the linear shift current conductivity is symmetric under interchange of the last two indices, $\sigma_{abc}^{\mathrm{sh,L}}=\sigma_{acb}^{\mathrm{sh,L}}$, whereas the circular injection current susceptibility is antisymmetric, $\eta_{abc}^{\mathrm{inj,C}}= -\eta_{acb}^{\mathrm{inj,C}}$.
We therefore discuss $\sigma_{xyz}^{\mathrm{sh,L}}$ and $\eta_{xyz}^{\mathrm{inj,C}}$, together with their quantum-geometric origins, and subsequently examine the relation between the CPGE trace and the topological charges of the multifold crossings below.

\subsection{Linear shift current}

\begin{figure}[t]
\centering
\includegraphics[width=\columnwidth]{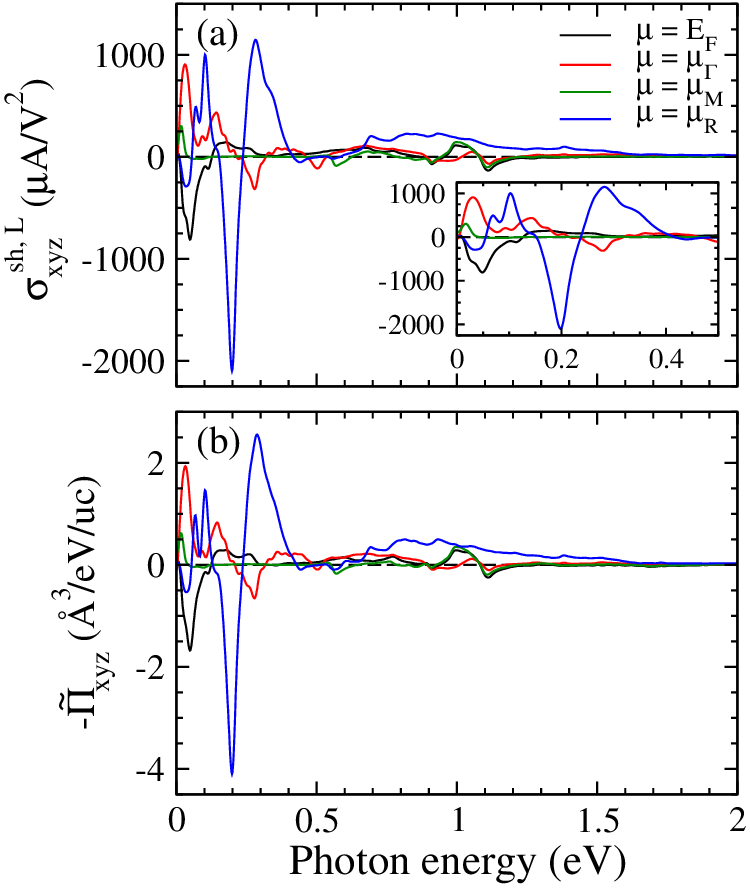}
\caption{(a) Linear shift current conductivity $\sigma_{xyz}^{\mathrm{sh,L}}$ of BeAu at the Fermi level and at chemical potentials aligned with the multifold crossings at $\Gamma$ ($\upmu_{\Gamma}$), M ($\upmu_{\mathrm{M}}$), and R ($\upmu_{\mathrm{R}}$). The inset shows the enlarged low-energy region. (b) Corresponding photon energy-resolved symplectic connection, $-\widetilde{\Pi}_{xyz}(\omega)$, for the four chemical potentials.}
\label{fig:shift}
\end{figure}

Figure~\ref{fig:shift}(a) shows the linear shift current conductivity $\sigma_{xyz}^{\mathrm{sh,L}}(\omega)$ at the Fermi level and at chemical potentials aligned with the multifold crossings at $\Gamma$, $M$, and $R$. 
At the Fermi level, the response decreases rapidly at low photon energies and reaches a pronounced negative maximum of $-810~\mu\mathrm{A/V^{2}}$ at $0.05$~eV. 
It then changes sign and attains a positive maximum of $144~\mu\mathrm{A/V^{2}}$ at $0.18$~eV. 
At higher photon energies, the conductivity is considerably smaller and exhibits a sequence of alternating positive and negative extrema. 

Shifting the chemical potential strongly modifies both the magnitude and sign of the linear shift current conductivity. 
For $\mu=\mu_{\Gamma}$, the conductivity exhibits a large positive low-energy peak of $907~\mu\mathrm{A/V^{2}}$ at $0.03$~eV. 
A second positive peak of $435~\mu\mathrm{A/V^{2}}$ occurs at $0.14$~eV, followed by a pronounced negative peak of $-314~\mu\mathrm{A/V^{2}}$ at $0.28$~eV. 
At $\mu=\mu_M$, a positive peak of $301~\mu\mathrm{A/V^{2}}$ appears at $0.02$~eV, whereas the response remains comparatively weak over much of the subsequent low- and intermediate-energy range. 
At higher photon energies, it develops alternating negative and positive features, including a positive peak of $145~\mu\mathrm{A/V^{2}}$ near $1.00$~eV. 
The largest response is obtained for $\mu=\mu_R$.
In this case, several large low-energy features of alternating sign appear, including a dominant negative peak of approximately $-2100~\mu\mathrm{A/V^{2}}$ at $0.20$~eV and a subsequent positive peak of $\sim1150~\mu\mathrm{A/V^{2}}$ at $0.28$~eV. 
The response then becomes substantially smaller at higher photon energies.

To better understand the features in the linear shift current conductivity spectra, and considering its direct relationship with the symplectic connection, we present the corresponding photon energy-resolved $-\widetilde{\Pi}_{xyz}(\omega)$ spectra in Fig.~\ref{fig:shift}(b) [see
Eq.~(\ref{eq:energy_resolved_symplectic_connection})]. 
At the Fermi level, the dominant negative peak in $-\widetilde{\Pi}_{xyz}(\omega)$ occurs at $0.05$~eV and corresponds to the large negative peak in $\sigma_{xyz}^{\mathrm{sh,L}}(\omega)$. The subsequent positive and negative peaks, including those near $0.18$, $1.00$, and $1.11$~eV, also occur at nearly the same photon energies in the two spectra. 
A similar correspondence is observed for the shifted chemical-potential cases. 
In particular, the two positive peaks and the subsequent negative peak for $\mu=\mu_{\Gamma}$, the prominent features near $0.02$, $0.57$, $1.00$, and $1.11$~eV for $\mu=\mu_M$, and the large alternating low-energy peaks for $\mu=\mu_R$ are all captured by the corresponding 
$-\widetilde{\Pi}_{xyz}(\omega)$ spectra. 
Thus, the symplectic connection explains the prominent peaks, dips, and sign changes in the calculated linear shift current conductivity spectra.

These results show that chemical-potential tuning redistributes the symplectic connection spectral weight and thereby strongly modifies both the magnitude and sign of the linear shift current conductivity spectra.
In particular, aligning the chemical potential with the multifold crossing at the $R$ point produces the largest response among the four cases, whereas the $M$-aligned case yields a comparatively weak response over much of the low- and intermediate-energy range, demonstrating the strong tunability of the shift current in BeAu.

\subsection{Circular injection current}

\begin{figure}[t]
\centering
\includegraphics[width=\columnwidth]{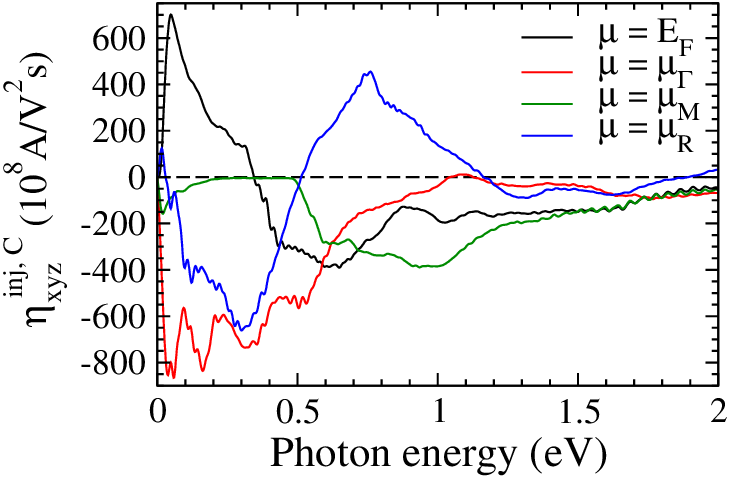}
\caption{Circular injection current susceptibility
$\eta_{xyz}^{\mathrm{inj,C}}$ of BeAu at the Fermi level and
at chemical potentials aligned with the multifold crossings at $\Gamma$ ($\upmu_{\Gamma}$), M ($\upmu_{\mathrm{M}}$), and R ($\upmu_{\mathrm{R}}$).}
\label{fig:injection}
\end{figure}

Figure~\ref{fig:injection} shows the circular injection current susceptibility $\eta_{xyz}^{\mathrm{inj,C}}(\omega)$ at the Fermi level and at chemical potentials aligned with the multifold crossings at $\Gamma$, $M$, and $R$. 
At the Fermi level, $\eta_{xyz}^{\mathrm{inj,C}}$ increases rapidly at low photon energies and reaches a pronounced positive maximum of $701\times10^{8}~\mathrm{A/(V^{2}s)}$ at $0.05$~eV. 
It then decreases, changes sign at approximately $0.35$~eV, and reaches a pronounced negative maximum of $-391\times10^{8}~\mathrm{A/(V^{2}s)}$ at $0.65$~eV. 
At higher photon energies, the response remains predominantly negative and gradually decreases in magnitude. 

The circular injection current susceptibility spectra change substantially upon aligning the chemical potential with the multifold crossings.
For $\mu=\mu_{\Gamma}$, the response is predominantly negative and reaches its dominant negative peak of $-865\times10^{8}~\mathrm{A/(V^{2}s)}$ at $0.06$~eV. 
Several additional negative features occur in the low-energy region, whereas its magnitude becomes substantially smaller at higher photon energies.
At $\mu=\mu_M$, $\eta_{xyz}^{\mathrm{inj,C}}$ remains negative throughout the calculated photon-energy range. 
The low-energy response is comparatively weak, while a negative peak of $-389\times10^{8}~\mathrm{A/(V^{2}s)}$ occurs at $0.94$~eV.
For $\mu=\mu_R$, $\eta_{xyz}^{\mathrm{inj,C}}$ 
reaches a pronounced negative peak of $-662\times10^{8}~\mathrm{A/(V^{2}s)}$ at $0.30$~eV. 
It then changes sign at approximately $0.51$~eV and attains a positive maximum of $455\times10^{8}~\mathrm{A/(V^{2}s)}$ at $0.76$~eV.
The response changes sign again at approximately $1.16$~eV and remains comparatively small at higher photon energies.
These results demonstrate the strong chemical-potential tunability of $\eta_{xyz}^{\mathrm{inj,C}}$ in BeAu, with the largest negative response obtained for $\mu=\mu_{\Gamma}$ and pronounced sign reversals appearing for $\mu=\mu_R$. 

Unlike the linear shift current conductivity, whose spectral features can be directly understood from the photon energy-resolved symplectic connection, the circular injection current susceptibility contains the band- and momentum-resolved product of the Berry curvature and group velocity difference. 
The photon energy-resolved Berry curvature $\Omega_{yz}(\omega)$ and group velocity difference
$\Delta^{x}(\omega)$ are obtained from separate band and Brillouin-zone integrations [see Eqs.~(\ref{eq:energy_resolved_berry_curvature}) and (\ref{eq:energy_resolved_group_velocity_difference}), respectively].
As a result, these separately integrated quantities do not preserve the transition-resolved correlation between the Berry curvature and group velocity difference that enters $\eta_{xyz}^{\mathrm{inj,C}}(\omega)$.
Consequently, neither spectrum alone nor the product of the two spectra is equivalent to the band- and momentum-resolved product of the group velocity difference and Berry curvature,
$\Delta_{mn}^{x}(\bm{k})\Omega_{yz}^{mn}(\bm{k})$, that determines $\eta_{xyz}^{\mathrm{inj,C}}$.

\subsection{CPGE trace and multifold topology}

To examine how multifold topology manifests in the circular injection current response, we consider the trace of the CPGE tensor. 
The second-rank CPGE tensor is related to the third-rank injection current susceptibility by contracting the two optical indices with the Levi-Civita tensor~\cite{deJuan2017},
\begin{equation}
\beta_{ad}(\omega)
=
\frac{1}{2}\varepsilon_{dbc}
\eta_{abc}^{\mathrm{inj}}(\omega),
\label{eq:beta_from_eta}
\end{equation}
where summation over the optical indices $b$ and $c$ is implied. 
Because $\varepsilon_{dbc}$ is antisymmetric under the interchange $b\leftrightarrow c$, this contraction selects the antisymmetric part of the injection current response that couples to circularly polarized light.
Using
$\eta_{acb}^{\mathrm{inj}}(\omega)
=
[\eta_{abc}^{\mathrm{inj}}(\omega)]^{*}$,
the diagonal components of the CPGE tensor can be written as
\begin{align}
\beta_{xx}(\omega)
&=
\frac{1}{2}
\left[
\eta_{xyz}^{\mathrm{inj}}(\omega)
-
\eta_{xzy}^{\mathrm{inj}}(\omega)
\right]
=
i\,\mathrm{Im}\,
\eta_{xyz}^{\mathrm{inj}}(\omega),
\\
\beta_{yy}(\omega)
&=
\frac{1}{2}
\left[
\eta_{yzx}^{\mathrm{inj}}(\omega)
-
\eta_{yxz}^{\mathrm{inj}}(\omega)
\right]
=
-i\,\mathrm{Im}\,
\eta_{yxz}^{\mathrm{inj}}(\omega),
\\
\beta_{zz}(\omega)
&=
\frac{1}{2}
\left[
\eta_{zxy}^{\mathrm{inj}}(\omega)
-
\eta_{zyx}^{\mathrm{inj}}(\omega)
\right]
=
i\,\mathrm{Im}\,
\eta_{zxy}^{\mathrm{inj}}(\omega).
\end{align}
Here, the factor of $1/2$ is cancelled by the factor of $2$ arising 
from the identity $\eta-\eta^{*}=2i\,\mathrm{Im}\,\eta$.
Defining the circular injection current susceptibility as
\begin{equation}
\eta_{abc}^{\mathrm{inj,C}}(\omega)
\equiv
\mathrm{Im}\,
\eta_{abc}^{\mathrm{inj}}(\omega),
\end{equation}
and using
$\mathrm{Tr}\,\beta=\beta_{xx}+\beta_{yy}+\beta_{zz}$,
we obtain
\begin{equation}
\frac{\mathrm{Tr}\,\beta(\omega)}{i}
=
\eta_{xyz}^{\mathrm{inj,C}}(\omega)
-
\eta_{yxz}^{\mathrm{inj,C}}(\omega)
+
\eta_{zxy}^{\mathrm{inj,C}}(\omega).
\label{eq:cpge_trace}
\end{equation}
We normalize the trace by the CPGE quantum $\beta_{0}=\pi e^{3}/h^{2}$.
In an ideal photon-energy window in which the optically activated momentum surfaces enclose isolated chiral multifold nodes, the normalized CPGE trace satisfies~\cite{deJuan2017,Flicker2018}
\begin{equation}
\frac{\mathrm{Tr}\,\beta(\omega)}{i\beta_{0}}=C,
\end{equation}
where $C$ is the total topological charge of the enclosed nodes.

\begin{figure}[t]
\centering
\includegraphics[width=\columnwidth]{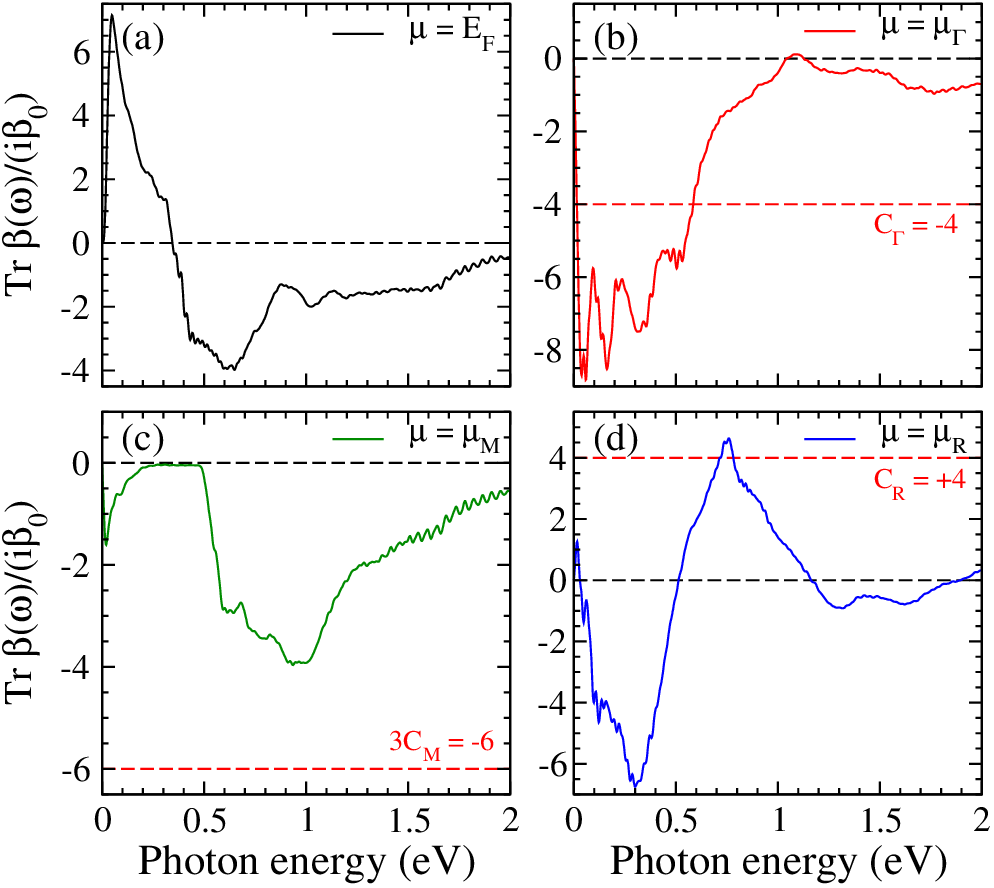}
\caption{Normalized CPGE trace
$\mathrm{Tr}\,\beta(\omega)/(i\beta_{0})$ of BeAu at (a) the Fermi level, and at chemical potentials aligned with the multifold crossings at (b) $\Gamma$ ($\upmu_{\Gamma}$), (c) M ($\upmu_{\mathrm{M}}$), and (d) R ($\upmu_{\mathrm{R}}$).
The horizontal dashed lines in panels (b)-(d) indicate the corresponding total Chern numbers $C_{\Gamma}=-4$, $3C_M=-6$, and $C_R=+4$, respectively.}
\label{fig:cpge}
\end{figure}

Figure~\ref{fig:cpge} shows the normalized CPGE trace at the Fermi level and at chemical potentials aligned with the multifold crossings at $\Gamma$, $M$, and $R$. 
The horizontal dashed lines in Figs.~\ref{fig:cpge}(b)--\ref{fig:cpge}(d) indicate the corresponding total Chern numbers, $C_{\Gamma}=-4$, $3C_M=-6$, and $C_R=+4$, respectively (see Appendix~\ref{app:topological_charges}).
For the $M$-aligned case, the factor of three accounts for the three symmetry-related $M$ points, each carrying $C_M=-2$.
At the Fermi level, the CPGE trace is strongly photon-energy dependent and reaches values larger in magnitude than the topological charge of any individual high-symmetry multifold crossing. 
This behavior may reflect the complex multiband Fermi surface of
BeAu~\cite{Vocaturo2026},  
suggesting that contributions beyond a single multifold node are relevant to the Fermi-level response.

For $\mu=\mu_{\Gamma}$, the CPGE trace remains predominantly negative and strongly overshoots $C_{\Gamma}=-4$ at low photon energies, crossing $-4$ only over a very narrow higher-energy interval where it continues to vary appreciably. 
The narrow crossing of the expected integer value does not constitute a quantized plateau and suggests that the optical response is not dominated solely by the fourfold node over this energy range.
For $\mu=\mu_M$, the low-energy trace remains negative but does not approach a stable value of $3C_M=-6$, and no quantized plateau is observed. 
The response for $\mu=\mu_R$ is qualitatively different: the trace changes from negative to positive with increasing photon energy and approaches $C_R=+4$ over a narrow energy interval before reaching a somewhat larger maximum. 
This sign reversal suggests competition between negative low-energy contributions and positive contributions that become dominant at higher photon energies.

These results distinguish a large chemical-potential-sensitive circular photocurrent from a quantized CPGE. 
The large circular injection current susceptibility in BeAu reflects strong Berry curvature and interband group velocity difference effects, whereas quantization additionally requires the optical response to be dominated by transitions associated with an optically isolated chiral node.
The complex multiband Fermi surface and competing interband transitions prevent this condition from being satisfied over an extended photon-energy range. 
BeAu therefore supports large and strongly chemical-potential-dependent circular photocurrents without exhibiting robust CPGE quantization.

\section{CONCLUSIONS}

In summary, we have systematically studied the linear optical conductivity and bulk photovoltaic responses of chiral BeAu using fully relativistic first-principles calculations.
We find that the calculated optical responses are large and strongly tunable by the chemical potential. 
The absorptive interband optical conductivity is quantitatively reproduced by $(e^{2}/\hbar)\omega g_{xx}(\omega)$, demonstrating that its spectral features originate from the variation of the photon energy-resolved quantum-metric spectral weight together with the explicit photon energy factor. 
Furthermore, the prominent peaks and sign changes in the linear shift current conductivity are explained by the corresponding symplectic connection spectra, while the circular injection current susceptibility is governed by the transition-resolved product of Berry curvature and the interband group velocity difference. 
At the Fermi level, BeAu exhibits a large low-energy linear shift current conductivity, reaching $-810~\mu\mathrm{A/V^2}$ at $0.05$ eV.
Aligning the chemical potential with the $R$-point crossing produces the largest linear shift current conductivity, whereas alignment with the three multifold crossings gives rise to distinct circular injection current susceptibility spectra.
Although the CPGE trace reflects the signs of the underlying topological charges, the complex multiband Fermi surface and competing interband transitions prevent a broad quantized plateau.
These findings demonstrate that BeAu supports large and strongly chemical-potential-dependent quantum-geometric optical responses while illustrating the distinction between enhanced circular photocurrents and robust CPGE quantization.
We hope that this study will stimulate further experimental investigations of the optical responses of chiral BeAu.

\begin{acknowledgments}
B.B.P. thanks Masahiro Fukuda for helpful discussions regarding the OpenMX code. 
B.B.P.’s current position is supported by the Institute for Solid State Physics (ISSP), The University of Tokyo.
\end{acknowledgments}

\appendix

\begin{figure}[t]
\centering
\includegraphics[width=\columnwidth]{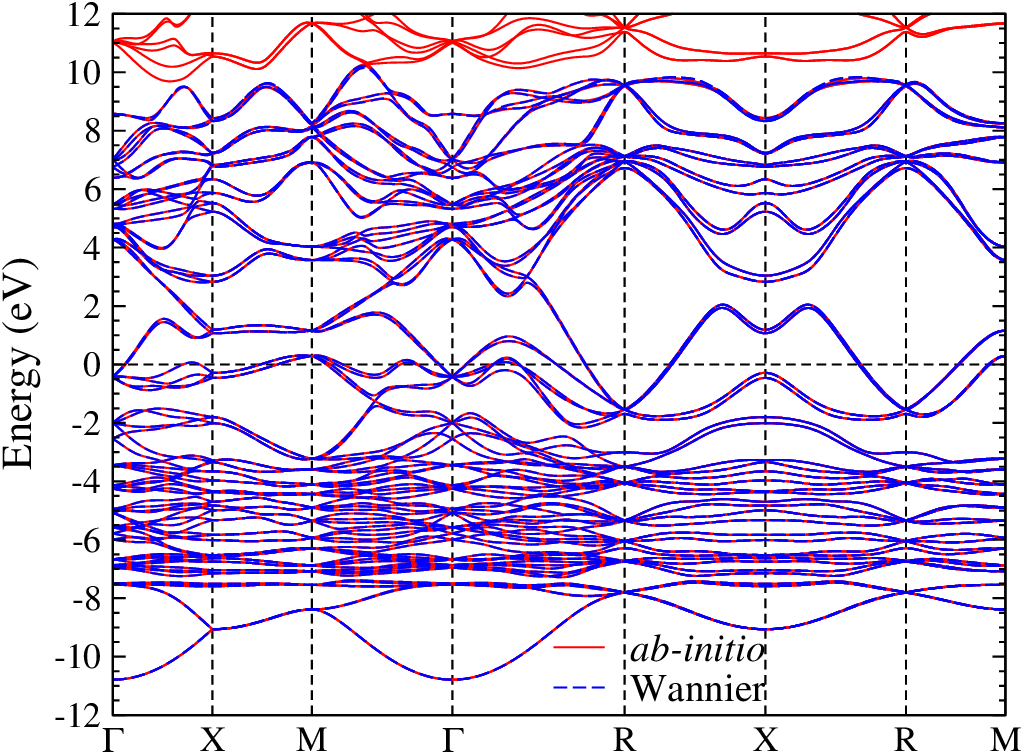}
\caption{Relativistic band structure (solid red lines) and the CWFs interpolated band structure (dashed blue lines) of BeAu. The horizontal dashed line denotes the Fermi level.}
\label{fig:wannier}
\end{figure}

\begin{figure*}[t]
\centering
\includegraphics[width=\textwidth]{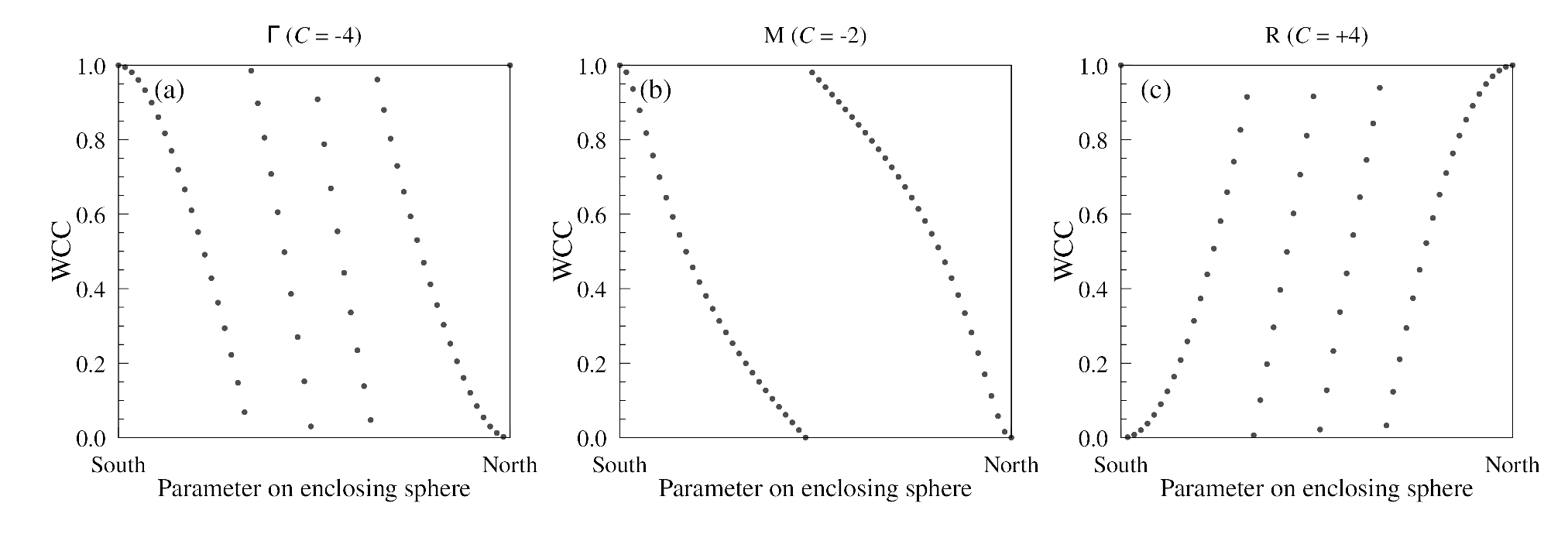}
\caption{Evolution of the hybrid Wannier charge centers (WCCs) on spherical surfaces enclosing the multifold crossings at (a) $\Gamma$, (b) $M$, and
(c) $R$, yielding the corresponding Chern numbers $C_{\Gamma}=-4$, $C_{M}=-2$, and $C_{R}=+4$, respectively.}
\label{fig:wcc}
\end{figure*}

\section{Relation between the optical conductivity and quantum metric}
\label{app:optical_metric}

In this Appendix, we derive the relation between the absorptive interband optical conductivity and the photon energy-resolved quantum metric. 
Starting from Eq.~(\ref{eq:optical_conductivity}), the diagonal component of the complex optical conductivity is
\begin{equation}
\sigma_{aa}(\omega)
=
\frac{i e^{2}}{\hbar}
\int\frac{d^{3}k}{(2\pi)^{3}}
\sum_{nm}
f_{mn}\,
\frac{
\omega_{mn}\,
r_{nm}^{a}(\mathbf{k})r_{mn}^{a}(\mathbf{k})
}{
\omega_{mn}-\omega-i\eta/\hbar
}.
\label{eq:app_optical_conductivity}
\end{equation}
In the limit $\eta\rightarrow0^{+}$, the denominator in Eq.~(\ref{eq:app_optical_conductivity}) can be treated using the Sokhotski--Plemelj identity,
\begin{equation}
\lim_{\eta\rightarrow0^{+}}
\frac{1}{x-i\eta/\hbar}
=
\mathcal{P}\frac{1}{x}
+i\pi\delta(x),
\label{eq:app_plemelj_identity}
\end{equation}
where $\mathcal{P}$ denotes the Cauchy principal value. 
Then, the real part of Eq.~(\ref{eq:app_optical_conductivity}) becomes
\begin{align}
\mathrm{Re}\,\sigma_{aa}(\omega)
={}&
-\frac{\pi e^{2}}{\hbar}
\int\frac{d^{3}k}{(2\pi)^{3}}
\sum_{nm}
f_{mn}\,
\omega_{mn}\,
r_{nm}^{a}(\mathbf{k})r_{mn}^{a}(\mathbf{k})
\nonumber\\
&\times
\delta(\omega_{mn}-\omega).
\label{eq:app_real_conductivity}
\end{align}
For a diagonal component, the quantum metric is
$g_{aa}^{mn}(\mathbf{k})
=\mathrm{Re}[r_{nm}^{a}(\mathbf{k})r_{mn}^{a}(\mathbf{k})]
=|r_{nm}^{a}(\mathbf{k})|^{2}$.
Using this relation together with $f_{nm}=-f_{mn}$,
Eq.~(\ref{eq:app_real_conductivity}) can be rewritten as
\begin{equation}
\mathrm{Re}\,\sigma_{aa}(\omega)
=
\frac{\pi e^{2}}{\hbar}
\int\frac{d^{3}k}{(2\pi)^{3}}
\sum_{nm}
f_{nm}\,
\omega_{mn}\,
g_{aa}^{mn}(\mathbf{k})
\delta(\omega_{mn}-\omega).
\label{eq:app_real_conductivity_metric}
\end{equation}
Here, the delta function imposes the resonance condition
$\omega_{mn}=\omega$. Using the photon energy-resolved quantum metric,
defined as
\begin{equation}
g_{aa}(\omega)
=
\pi
\int\frac{d^{3}k}{(2\pi)^{3}}
\sum_{nm}
f_{nm}\,
g_{aa}^{mn}(\mathbf{k})
\delta(\omega_{mn}-\omega),
\label{eq:app_energy_resolved_metric}
\end{equation}
Eq.~(\ref{eq:app_real_conductivity_metric}) then gives
\begin{equation}
\boxed{
\mathrm{Re}\,\sigma_{aa}(\omega)
=
\frac{e^{2}\omega}{\hbar}\,
g_{aa}(\omega)
}.
\label{eq:app_sigma_metric_relation}
\end{equation}
Thus, apart from the prefactor $e^{2}/\hbar$, the frequency dependence of the absorptive optical conductivity is determined by $\omega g_{aa}(\omega)$. 

\section{Comparison between the first-principles and CWF band structures}
\label{app:cwf_bands}

Figure~\ref{fig:wannier} compares the fully relativistic first-principles band structure of BeAu with that obtained from the closest Wannier function (CWF) Hamiltonian along the high-symmetry path. 
The electronic structure exhibits several multifold band crossings characteristic of the chiral cubic space group $P2_{1}3$. 
In particular, fourfold crossings occur at the $\Gamma$ and $M$ points, while a sixfold crossing is present at the $R$ point. 
Relative to the Fermi level, these crossings are located at approximately $-0.404$, $+0.288$, and $-1.542$~eV, respectively. 
The Fermi level also intersects additional dispersive bands, indicating that the low-energy electronic structure cannot be described by an isolated multifold node alone.

As shown in Fig.~\ref{fig:wannier}, the CWF-interpolated bands closely reproduce the first-principles band structure throughout the energy range relevant to the optical response calculations. 
In particular, the locations and dispersions of the bands forming the multifold crossings at $\Gamma$, $M$, and $R$ are accurately retained. 
This agreement supports the use of the CWF Hamiltonian for dense $k$-point interpolation of the optical matrix elements and response functions presented in the main text.

\section{Topological charges of the multifold nodes}
\label{app:topological_charges}

The topological charges of the multifold crossings are determined from the evolution of the hybrid Wannier centers on spherical surfaces enclosing the individual nodes. 
The Chern number is obtained from the net winding of the hybrid Wannier centers as the enclosing surface is traversed.

Figure~\ref{fig:wcc} shows the hybrid Wannier center evolution for the multifold crossings at $\Gamma$, $M$, and $R$. 
The corresponding windings yield $C_{\Gamma}=-4$, $C_{M}=-2$, and $C_{R}=+4$, respectively. 
Here, $C_M=-2$ denotes the Chern number of an individual $M$-point crossing; the three symmetry-related $M$ points therefore carry a total Chern number of $3C_M=-6$.
These results confirm the topological character of the fourfold crossings at $\Gamma$ and $M$ and the sixfold crossing at $R$, and are used to interpret the CPGE spectra discussed in the main text.

\end{document}